\documentclass{ws-ijmpd}

\begin{document}

\markboth{Zhang et al.} {Spectral Variability of Individual TeV Blazars}

%
\catchline{}{}{}{}{}
%

\title{BLAZAR ANTI-SEQUENCE OF SPECTRAL VARIABILITY FOR INDIVIDUAL TeV BLAZARS}

\author{Jin Zhang$^{1,2}$, Shang-Nan Zhang$^{3,1}$, En-Wei Liang$^{4}$}

\address{$^{1}$National Astronomical Observatories, Chinese Academy of Sciences, Beijing, 100012, China; zhang.jin@hotmail.com\\
$^{2}$College of Physics and Electronic Engineering, Guangxi Teachers Education University, Nanning, 530001, China\\
$^{3}$Key Laboratory of Particle Astrophysics, Institute of High Energy Physics, Chinese Academy of Sciences, Beijing 100049, China\\
$^{4}$Department of Physics and GXU-NAOC Center for Astrophysics and Space Sciences, Guangxi University,
Nanning, 530004, China\\} \maketitle

\begin{history}
\received{Day Month Year} \revised{Day Month Year} \comby{Managing Editor}
\end{history}

\begin{abstract}
We compile from literature the broadband SEDs of twelve TeV blazars observed simultaneously or
quasi-simultaneously with {\em Fermi}/LAT and other instruments. Two SEDs are available for each of the objects
and the state is identified as a low or high state according to its flux density at GeV/TeV band. The observed
SEDs of BL Lac objects (BL Lacs) are fitted well with the synchrotron + synchrotron-self-Compton (syn+SSC) model,
whereas the SEDs of the two flat spectrum radio quasars (FSRQs) need to include the contributions of external
Compton scattering. In this scenario, it is found that the Doppler factor $\delta$ of FSRQs is smaller than that
of BL Lacs, but the magnetic field strength $B$ of FSRQs is larger than that of BL Lacs. The increase of the
peak frequency of the SEDs is accompanied with the increase of the flux for the individual sources, which seems
opposite to the observational phenomena of the blazar sequence. We refer this phenomenon to {\em blazar
anti-sequence} of spectral variability for individual TeV blazars. However, both the blazar sequence from FSRQs
to BL Lacs and blazar anti-sequence of the spectral variability from low state to high state are accompanied by
an increase of the break Lorentz factor of the electron's spectrum $\gamma_{\rm b}$ and a decrease of $B$. We
propose a model in which the mass accretion rate $\dot M$ is the driving force behind both the blazar
sequence for ensembles of blazars and the blazar anti-sequence for individual blazars. Specifically we suggest
that the differences in $\langle\dot M\rangle$ of different blazars produce the observed blazar sequence, but
$\Delta\dot M$ in each blazar results in the observed blazar anti-sequence.

\end{abstract}

\keywords{Radiation mechanisms: non-thermal; BL Lacertae objects: general; quasars: general; gamma-rays: theory}

\section{Introduction}

Blazars, a subsample of active galactic nuclei (AGNs), are composed of BL Lac objects (BL Lacs) and flat
spectrum radio quasars (FSRQs). It is well known that the observed emission from blazars is jet dominated and
the observed broadband spectral energy distributions (SEDs) are bimodal. Generally, the bump at the
IR-optical-UV band is explained with the synchrotron process of relativistic electrons accelerated in the jets
and the bump at the GeV-TeV gamma-ray band is due to the inverse Compton (IC) scattering of the same electron
population. The seed photon fields may be from the synchrotron radiation themselves (the so-called SSC
model)\cite{Maraschi1992,Ghisellini1996} or from external radiation fields (EC), such as the broad line region
(BLR)\cite{Dermer1992}. Broadband SEDs obtained simultaneously or quasi-simultaneously are critical to
investigate the radiation mechanisms and the physical properties of the emitting regions for the two kinds of
blazars.

With a large sample of different types of blazars, Fossati et al. (1998) reported a spectral sequence of
FSRQ--LBL--HBL\cite{Fossati1998}, i.e., along with this sequence, observationally, an increase in the peak
frequency of synchrotron radiation ($\nu_{\rm s}$) corresponds to the decreases of bolometric luminosity and the
ratio ($L_{\rm IC}/L_{\rm syn}$) of the luminosities for the high- and low-energy spectral components.
Theoretically, an anti-correlation between the break Lorentz factor of electrons ($\gamma_{\rm b}$) and the
energy density of radiation regions ($U^{'}$) in the comoving frame was found. So this blazar sequence was
interpreted by Ghisellini et al. (1998) with the cooling of the external photon fields (such as the BLRs) of the
blazars\cite{Ghisellini1998}. The discoveries of `blue quasars' posed a challenge to the previous interpretation
for blazar sequence. To address this problem, Ghisellini \& Tavecchio (2008) reported that, more physically, the
sequence is due to the different black hole (BH) masses and accretion rates of the sources\cite{Ghisellini2008}.
According to this interpretation, the `blue quasars' should have large BH masses and intermediate accretion
rates and thus the emission region is beyond the BLR. Moreover, the red low-luminosity blazars should exist,
which have small BH masses and relatively large accretion rates. More recently, Chen \& Bai (2011) extended this
sequence to narrow line Seyfert 1 galaxies\cite{Chen2011}, which are similar to the low-peak-frequency and
low-luminosity blazars.

Observations show that the SED peak frequency of a TeV blazar increases with its flux, indicating a tendency
that a brighter TeV emission corresponds to a harder spectrum for the emission at the X-ray and gamma-ray
bands\cite{Anderhub2009,Zhang2011}. This phenomenon seems to be opposite to the blazar sequence; we thus refer
this phenomenon to {\em blazar anti-sequence} of spectral variability for individual TeV blazars. The spectral
shift at different states of TeV emission for the individual sources may be due to the different $\gamma_{\rm
b}$ at different states\cite{Zhang2011}. Thus, it seems that both the blazar sequence and anti-sequence are
caused by the change of $\gamma_{b}$. However it remains unclear whether there are any connections between the
two phenomena. So far, about forty AGNs have been detected at the TeV gamma-ray band. Except for two radio
galaxies, all of the confirmed TeV AGNs are blazars. Moreover, most of the confirmed sources by {\em Fermi}/LAT
are also blazars. The abundant observation data provide an excellent opportunity to reveal the physical
connections between the blazar sequence and anti-sequence.

\section{Sample Selection}

In order to obtain well-sampled SEDs, only the TeV blazars, which have positive {\em Fermi}/LAT detections and
two or more observed broadband SEDs, are considered. Twelve blazars are included in our sample, two FSRQs (3C
279 and PKS 1510-089) and ten BL Lacs (W Com, Mkn 421, Mkn 501, PKS 2155-304, 1ES 1101-232, BL Lacertae, 1ES
2344+514, 1ES 1959+650, PKS 2005-489, S5 0716+714). We compile their broadband SEDs that were simultaneously or
quasi-simultaneously observed with {\em Fermi}/LAT and other instruments from literature. The two
well-sampled SEDs for a source are identified as a low or high state according to the observed or extrapolated
flux density at 1 TeV, except for PKS 1510-089. Because the SEDs of PKS 1510-089 are cut off at 1 TeV, the high
and low states of this source are defined by the flux density at 10 GeV. The observation data of the broadband
SEDs for 3C 279 and PKS 1510-089 are from literature\cite{Abdo2010a,Abdo2010b,Kataoka2008}.

\section{Models and Results}

The observed SEDs for those sources are double-peaked. For BL Lacs the external photon fields from their BLRs,
if exist, are very weak and negligible, compared with the synchrotron radiation photon fields. Therefore, only
the syn+SSC model is considered to fit the observed broadband SEDs of the ten BL Lacs. The detailed calculation
process for the model and the strategy for parameter constraints can be found in Ref.\cite{Zhang2011}. Following
the methods described in Ref.\cite{Zhang2011}, the SEDs for the ten BL Lacs are fitted well by the single-zone
syn+SSC model; the fitting results and parameters can be found in Fig.~1 and table 1 of Ref.\cite{Zhang2011}.

For the two FSRQs, the contributions of external field photons from their BLRs need to be considered. The total
luminosities from the BLRs for the two FSRQs are taken from Ref.\cite{Celotti1997} and then the BLR sizes are
calculated using the BLR luminosity with formula (23) given in Ref.\cite{Liu2006}. The radiation from a BLR is
assumed to be a blackbody spectrum and the corresponding energy densities seen in the comoving frame are
$U^{'}_{\rm BLR}=5.37\times10^{-3}\Gamma^{2}$ erg cm$^{-3}$ and $U^{'}_{\rm BLR}=4.98\times10^{-3}\Gamma^{2}$
erg cm$^{-3}$ for 3C 279 and PKS 1510-089, respectively, where we take $\Gamma=\delta$. Thus, the syn+SSC+EC/BLR
model is used to fit the observed SEDs of the two FSRQs and the model can explain the observed SEDs well as
shown in Fig.~1.

The range of $\delta$ for these BL Lacs is from 8 to 50 and is clustered at 11 $\sim$ 30. On average, $\delta$ of BL Lacs is larger than that of the FSRQs, but the magnetic field strength $B$ of BL Lacs is smaller than that of the FSRQs, consistent with the
results of some previous works\cite{Ghisellini2010,Celotti2008}. The values of $\gamma_{\rm b}$ for these BL
Lacs vary from $10^3$ to $10^6$, but the values of $B$ are clustered around $0.1 \sim 0.6$ G. No correlation
between $\gamma_{\rm b}$ and $B$ for these BL Lacs is found.

\section{The Blazar Anti-Sequence and the Blazar Sequence}
\subsection{The Blazar Anti-Sequence}

As shown in Fig.~1 and Fig.~1 in Ref.\cite{Zhang2011}, there are clear spectral shifts accompanying the flux
variations for these TeV blazars. The SEDs of the twelve TeV blazars in our sample are obtained in the low and
high TeV states, which is defined according to the flux density at 1 TeV except for PKS 1510-089, for which the
state is defined according to its flux density at 10 GeV. We compare $\nu_{\rm s}$ and $\nu_{\rm c}$ between the
high and low states in Fig.~2(a). It is clear that the SEDs in the high states shift to a higher energy band
than that in the low state.

As reported in our previous paper\cite{Zhang2011}, to investigate what may be responsible for the spectral shift
in the low and high states, we derive the ratios of the flux density at 1 TeV $ (R_{\rm 1\ TeV})$ and the
physical parameters ($R_x$) in the high state to that in the low state for these sources, where $x$ stands for
$L_{\rm bol}$, $B$, $\delta$, $\gamma_{\rm b}$, or $P_{\rm jet}$. The results on the SEDs of PKS 1510-089 are
not included here. The values of $\gamma_{\rm b}$, $L_{\rm bol}$, and $P_{\rm jet}$ of the high states are
systematically higher than that of the low state. A tentative correlation between $R_{\gamma_{\rm b}}$ and
$R_{\rm 1\ TeV}$ is found with a correlation coefficient $r=0.55$ and a chance probability $p=0.077$ as shown in
Fig.~2(b). Therefore, it is possible that the spectral shift at different states is due to the different
$\gamma_{\rm b}$ of these sources.

In order to compare the differences of parameters for the high and low states, we also calculate the magnetic
field energy density ($U_{B}$), the available photon energy density ($U^{'}_{\rm ph}$, $U^{'}_{\rm
ph}=U^{'}_{\rm BLR}+U^{'}_{\rm syn, avail}$ for FSRQs and $U^{'}_{\rm ph}=U^{'}_{\rm syn, avail}$ for BL Lacs)
for IC process in the comoving frame, the luminosities of the synchrotron radiation ($L_{\rm syn}$) and the IC
process ($L_{\rm IC}$, $L_{\rm IC}=L_{\rm SSC}$ for BL Lacs and $L_{\rm IC}=L_{\rm SSC}+L_{\rm EC}$ for FSRQs)
in high and low states. The ratios of those parameters for the two states ($R_{U_{B}}$, $R_{U^{'}_{\rm ph}}$,
$R_{L_{\rm syn}}$, and $R_{L_{\rm IC}}$) as a function of the ratio of $\gamma_{\rm b}$ ($R_{\gamma_{\rm b}}$)
are shown in Fig.~3. No correlations between the ratios of those parameters and the ratio of $\gamma_{\rm b}$
are found. As presented in Fig.~3(a), for most
of the sources $U_{B}$ becomes smaller in the high state than in the low state. However, the ratios of $U^{'}_{\rm ph}$, $L_{\rm syn}$, and $L_{\rm IC}$ are larger than unity for most of the
sources as presented in Fig.~3(a), (b).

\subsection{The Blazar Sequence}
As described in section 1, according to the observational phenomenon of the blazar
sequence\cite{Ghisellini1998}, one can expect that both $L_{\rm bol}$ and $L_{\rm IC}/L_{\rm syn}$ are
anti-correlated with $\nu_{\rm s}$. $L_{\rm bol}$ and $L_{\rm IC}/L_{\rm syn}$ as a function of $\nu_{\rm s}$,
are shown in Fig.~4(a) and (b). A weak correlation is found for $\nu_{\rm s} - L_{\rm bol}$. The Spearman
correlation analysis yields a correlation coefficient $r=-0.58$ and a chance probability $p=0.05$ for the high
state data, $r=-0.61$ and $p=0.04$ for low state data, respectively. Excluding the two FSRQs, no correlation
would be found. However, the ratio of $L_{\rm IC}/L_{\rm s}$ is indeed anti-correlated with $\nu_{\rm s}$,
especially in their low states. The Spearman correlation analysis yields a correlation coefficient $r=-0.71$ and
a chance probability $p=0.009$ for the high state data, $r=-0.94$ and $p<10^{-4}$ for the low state data,
respectively. Because most of the sources in our sample are BL Lacs, the external photon fields outside their
jets are much weaker than the synchrotron radiation photon field and the EC process is thus not considered for
these sources. As $\nu_{\rm s}$ increases, the SEDs shift to the higher frequency end and the KN effect should
be more significant. According to equation (20) in Ref.\cite{Tavecchio1998}, $U^{'}_{\rm syn, avail}=U^{'}_{\rm
syn}(\frac{3mc^{2}\delta}{4h\gamma_{\rm b}\nu_{\rm s}})^{1-\alpha_{1}}$, the available photon energy density of
synchrotron radiation for IC process decreases with the increase of $\gamma_{\rm b}$.
For $\frac{L_{\rm SSC}}{L_{\rm syn}}\sim\frac{U^{'}_{\rm syn, avail}}{U_{B}}$, the
ratio of $L_{\rm SSC}/L_{\rm syn}$ would decrease as $\nu_{\rm s}$ increases, since $U_B$ is almost constant and
$U^{'}_{\rm syn, avail}$ decrease along with the increase of $\gamma_{\rm b}$ for the BL Lacs in our sample.
Therefore, the anti-correlation of $L_{\rm IC}/L_{\rm syn}-\nu_{\rm s}$ may be also due to the KN effect,
especially for BL Lacs.

Because the interpretation for blazar sequence is cooling of the external photon fields, a more ``theoretical"
scenario than the purely phenomenological sequence is the anti-correlation between the break Lorentz factor of
electrons $\gamma_{\rm b}$ and the energy density of radiation regions $U^{'}$ in the comoving
frame\cite{Ghisellini1998,Ghisellini2010}. $\gamma_{\rm b}$ as a function of $U^{'}$ ($U^{'}=U_{\rm
syn,avail}^{'}+U^{'}_{B}$ for BL Lacs, $U^{'}=U_{\rm syn,avail}^{'}+U_{\rm BLR}^{'}+U^{'}_{B}$ for FSRQs) is
also shown in Fig.~4(c). Although the correlation between $\gamma_{\rm b}$ and $U^{'}$ of our sample sources has
large scatters, especially for BL Lacs, comparing our results with that in Ref.\cite{Ghisellini2010}, the two
results are consistent and the BL Lacs included in our sample distribute in the left top of Fig.~4(c), where the
EC process is not important. If the different $\gamma_{\rm b}$ is totally due to the different
external photon field, there should be a correlation between the luminosity of the BLR and the peak frequency of
synchrotron emission. With a FSRQs sample\cite{Chen2009}, however, no correlation between $L_{\rm BLR}$ and
$\nu_{\rm s}$ is found as shown in Fig.~5(a). Sometimes, for simplicity, the luminosity and the radius of a BLR
is assumed as $L_{\rm BLR}=0.1L_{\rm disk}$ and $R_{\rm BLR}=10^{17}L_{\rm disk,45}^{1/2}$. So the comoving
energy density of a BLR is given by $U^{'}_{\rm BLR}=3.76\times10^{-2}\Gamma^{2}$ erg cm$^{-3}$ under these
assumptions\cite{Celotti2008} and is totally decided by the value of $\Gamma=\delta$. Nevertheless, no
correlation between $\Gamma^{2}$ and $\gamma_{\rm b}$ is found either, as presented in Fig.~5(b). $\gamma_{\rm
b}$ is correlated with the total energy density of the emitting regions in the comoving frame as shown in
Fig.~4(c), but not correlated with the energy density of the BLR. Using the sample data in Ref.\cite{Celotti2008} and our sample data, it is found that $\gamma_{\rm b}$ is correlated with the magnetic field energy density,
as shown in Fig.~5(c). Some works indeed demonstrate that the magnetic field strength $B$ of FSRQs is different
from and larger than that of BL Lacs\cite{Ghisellini2010,Celotti2008}, also consistent with our results. So the
blazar sequence may be due to the different $\gamma_{\rm b}$ and $B$ of these sources.

\subsection{Implications for Blazar Sequence and Anti-Sequence}

As shown in Fig.~1 of this paper and Fig.~1 of Ref.\cite{Zhang2011}, the characteristics of spectral evolution
at different states for a given source are opposite to the blazar sequence. In order to investigate the physical
connections of the two phenomena, firstly, we need to define another physical parameter ($R_{\rm Y}$), the ratio
of physical parameters in the high state to that in the low state for the twelve sources, where Y stands for
$\nu_{\rm s}$, $L_{\rm bol}$, $L_{\rm IC}/L_{\rm syn}$, $\gamma_{\rm b}$, and $U^{'}$. Comparing the spectral
evolution for a given source with the blazar sequence, we find the following:

\begin{romanlist}[(ii)]
\item According to the blazar sequence, the peak frequency of synchrotron radiation increases with decreasing bolometric luminosity.
The ratio of bolometric luminosity ($R_{L_{\rm bol}}$) as a function of the ratio of peak frequency
($R_{\nu_{\rm s}}$) for the two states is presented in Fig.~6(a). It is found that for most of the sources in
our sample the peak frequencies of synchrotron radiation move to the higher energy band in the high state than
in the low state; at the same time the bolometric luminosities increase.

\item According to the blazar sequence, the peak frequency of synchrotron radiation increases with decreasing ratio of the luminosities
for the IC and synchrotron components. However, for most of the sources in our sample the values of $L_{\rm
IC}/L_{\rm syn}$ for high state become larger than that for low state, accompanied with the increase of
synchrotron radiation peak frequency as shown in Fig.~6(b).

\item A more ``theoretical" characteristic for the blazar sequence is the anti-correlation between $\gamma_{\rm b}$ and $U^{'}$ in the comoving frame. However, as shown in Fig.~6(c), both $\gamma_{\rm b}$ and $U^{'}$ of the source in the high state become larger than that in the low state.

\end{romanlist}

As described above, both the blazar sequence from FSRQs to BL Lacs and anti-sequence of the individual sources
from low state to high state are accompanied by an increase of $\gamma_{\rm b}$ and a decrease of $B$, as shown
in Fig.~3(a) and Fig.~5(c). Ghisellini \& Tavecchio (2010) reported that the blazar sequence is linked to the
different BH masses and accretion rates of different sources\cite{Ghisellini2010}. We propose here that the
different states of the individual objects are also linked to the variations of the accretion rate for each source.
So both the blazar sequence (change of source type) and the blazar anti-sequence of spectral variability of
the individual objects (change of source state) are linked to the change of accretion rate. The flow chart
illustrating how the accretion rate drives the blazar sequence and the anti-sequence is shown in Fig.~7, which is
explained as follows:

\begin{itemlist}
\item Assuming the mass accretion rate $\dot M$ decreases, we can explain the blazar sequence (the left branch, {\em (a-d)} in the following) from FSRQs to BL Lacs and the blazar anti-sequence (the right branch, {\em (e-h)} in the following) for the individual sources from the low to high states.
\item {\em (a)---} Assuming equipartition between the magnetic energy and other forms of energies in the accretion disk and that the magnetic fields in the jets of blazars are carried over from the disk, then we expect $B$ decreases. If $\gamma_{\rm b}$ is determined by radiative cooling, then we expect $\gamma_{\rm b}$ increases.
\item {\em (b)---} Statistically AGNs with smaller BH masses have lower $\dot M$ (in physical units) and thus lower bolometric luminosity $L_{\rm bol}$.
\item {\em (c)---} The BLR exists only above a critical value of the accretion rate\cite{Ghisellini2008} and its luminosity is well correlated with $L_{\rm bol}$, though with a delay. It is well known that the emission lines of BL Lacs are very weak, thus the second bump of their SEDs are not contributed by IC/BLR ($L_{\rm EC}$). Therefore, the values of $L_{\rm IC}/L_{\rm syn}$ for FSRQs are larger that of BL Lacs as presented in Fig.~4(b).
\item {\em (d)---} From FSRQs to BL Lacs, $U_{\rm BLR}^{'}$ and $U_{B}$ (see Fig.~45(c)) decrease with the increasing $\gamma_{\rm b}$, resulting in an anti-correlation between $U^{'}$ and $\gamma_{\rm b}$ shown in Fig.~4(c).
\item {\em (e)---} {\em For an individual source from the low to high state ($\dot M$ decreases)}, the electron energy $\gamma_{\rm e}$ should increase with $\gamma_{\rm b}$ (see Fig.~2(b) for larger $\gamma_{\rm b}$ in high states) and thus the bolometric luminosity increases as shown in Fig.~6(a). {\em This explains (i) above.}
\item {\em (f)---} $U_B$ decreases with $\dot M$ (as discussed above and see Fig.~3(a)), but $U_{\rm syn}^{'}$ increases with $\gamma_{\rm b}$ (for increasing $L_{\rm syn}$).
\item {\em (g)---} $R_{L_{\rm IC}/L_{\rm syn}}$ increases for the BL Lacs in the high state as shown in Fig.~6(b), because $L_{\rm IC}$ is ``boosted" from $L_{\rm syn}$ by $\gamma_{\rm b}$. For the two FSRQs, $\delta$ is larger and $B$ is smaller in high state than that in the low state, corresponding to larger $U_{\rm BLR}^{'}$ and smaller $U_{B}$, so the ratio, $R_{L_{\rm IC}/L_{\rm syn}}$, is also larger than unity, as shown in Fig.~6(b). {\em This explains (ii) above.}
\item {\em (h)---} Less change of $U_{B}$ and more increase of $U_{\rm syn}^{'}$ result in $R_{U^{'}}$ larger than unity for most of the BL Lacs, as shown in Fig.~6(c). {\em This explains (iii) above.}
\end{itemlist}

\section{Summary}

We have compiled the broadband SEDs of twelve TeV blazars that were simultaneously or quasi-simultaneously
observed with {\em Fermi}/LAT and other instruments from literature. Each of those sources has two broadband
SEDs available, which are identified as a high or a low state according to its flux density at GeV/TeV band. We
found that the syn+SSC model can well represent the observed SEDs for BL Lacs, whereas the EC/BLR contribution
needs to be considered for explaining the observed SEDs of the two FSRQs. The magnetic field strength $B$ of the
two FSRQs is larger, but their Doppler factor $\delta$ is smaller than that of the ten BL Lacs. Significant
spectral shift to high energies accompanying with the flux increase is observed for each individual source,
which seems opposite to the observational phenomenon of the blazar sequence. We refer this phenomenon to the blazar
anti-sequence. However, it is found that both the blazar sequence from FSRQs to BL Lacs and the anti-sequence of
the individual sources from low to high states are accompanied by an increase of $\gamma_{\rm b}$ and a
decrease of $B$. We propose here that the blazar sequence and the anti-sequence of spectral variability of the
individual sources are driven by decreasing accretion rate. A simple flow chart, which describes qualitatively
how the accretion rate contributes to the blazar sequence and anti-sequence, is also given in this paper.

\section*{Acknowledgments}

This work is supported by the National Natural Science Foundation of China (Grants 11078008, 11025313, 10873002,
11133002, 10821061, 10725313), the National Basic Research Program (973 Programme) of China (Grant
2009CB824800), China Postdoctoral Science Foundation, Guangxi Science Foundation (2011GXNSFB018063,
2010GXNSFC013011), and Guangxi SHI-BAI-QIAN project (Grant 2007201).

\begin{figure*}
\includegraphics[angle=0,scale=0.27]{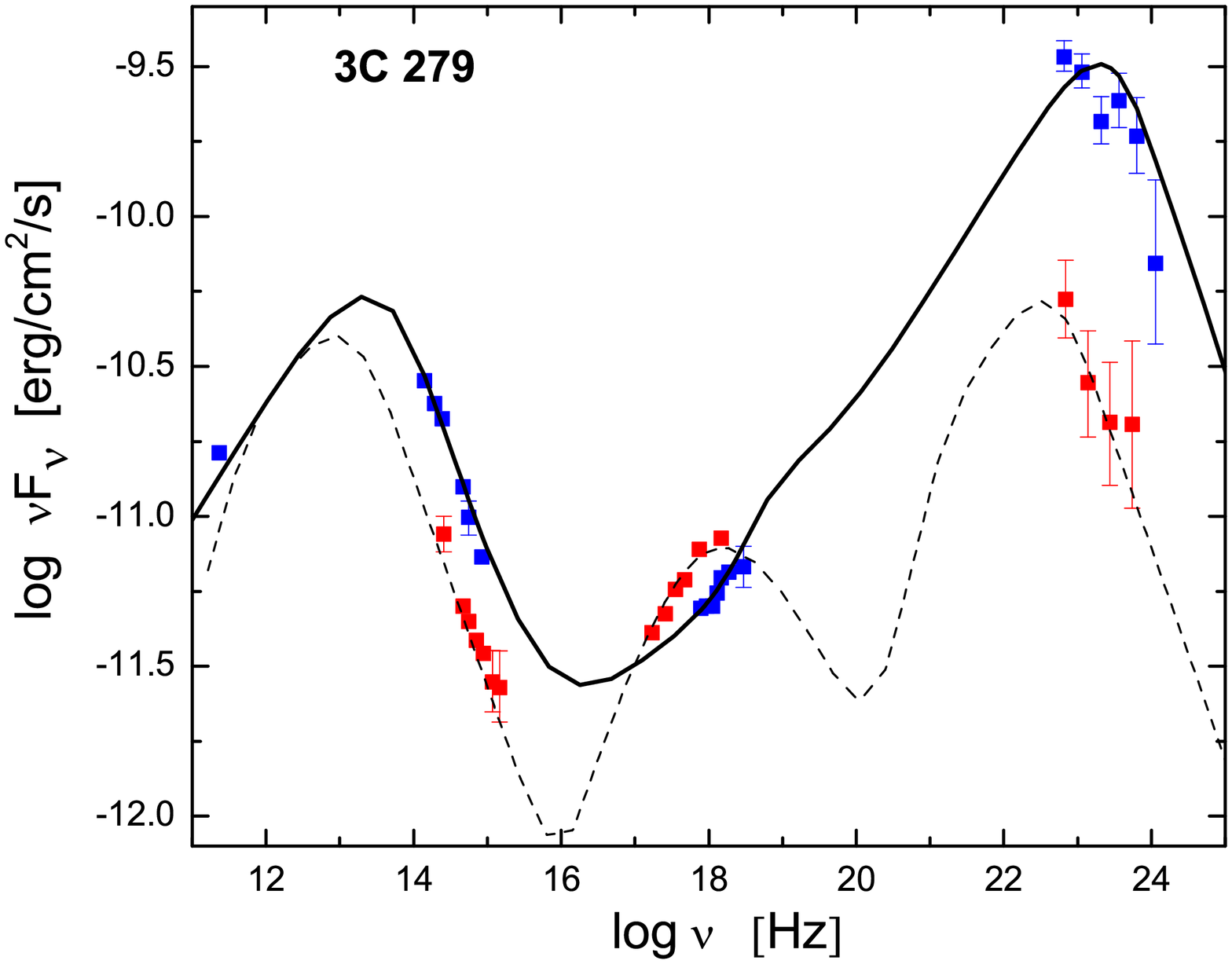}
\hfill
\includegraphics[angle=0,scale=0.27]{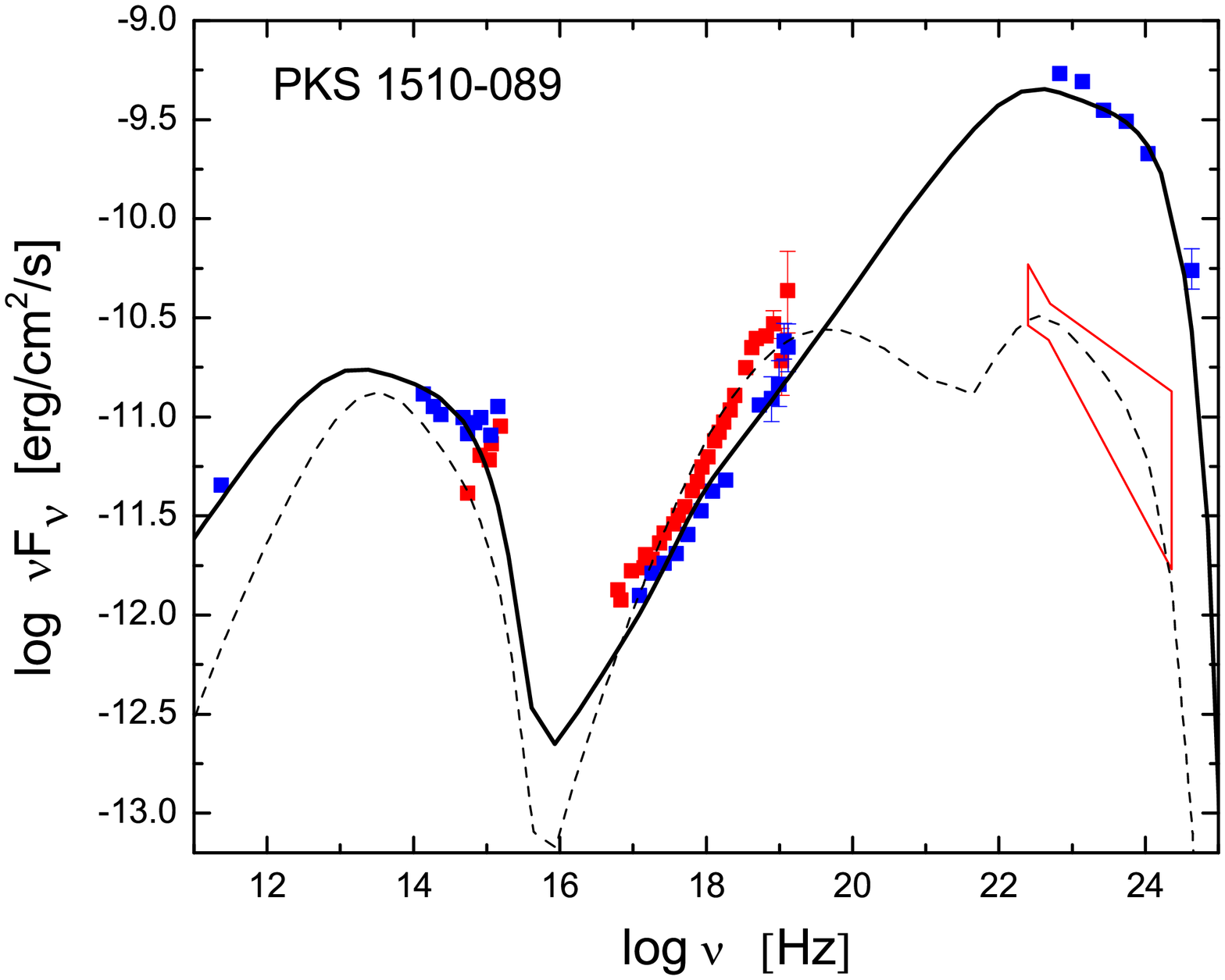}
\caption{The observed SEDs with model fitting for 3C 279 and PKS 1510-089. The data of high and low states are
marked with {\em blue and red} symbols, respectively. The fitting parameters of 3C 279 are $p_{1}=2.1$,
$p_{2}=4.4$, $\gamma_{\rm b}=294$, $\Delta t=24$ h, $B=4.$ G, $\delta=12$ for the low state SED and $p_{1}=2.2$, $p_{2}=4.46$, $\gamma_{\rm b}=526$, $\Delta t=24$ h, $B=2.65$ G, $\delta=17$ for the high state SED. The fitting parameters of PKS 1510-089 are $p_{1}=1.1$,
$p_{2}=3.8$, $\gamma_{\rm b}=631$, $\Delta t=12$ h, $B=2.1$ G, $\delta=8.8$ for the low state SED and $p_{1}=1.9$, $p_{2}=3.2$,  $\gamma_{\rm b}=400$, $\Delta t=24$ h, $B=1.15$ G, $\delta=16$ for the high state SED.}
\end{figure*}

\begin{figure*}
\includegraphics[angle=0,scale=0.27]{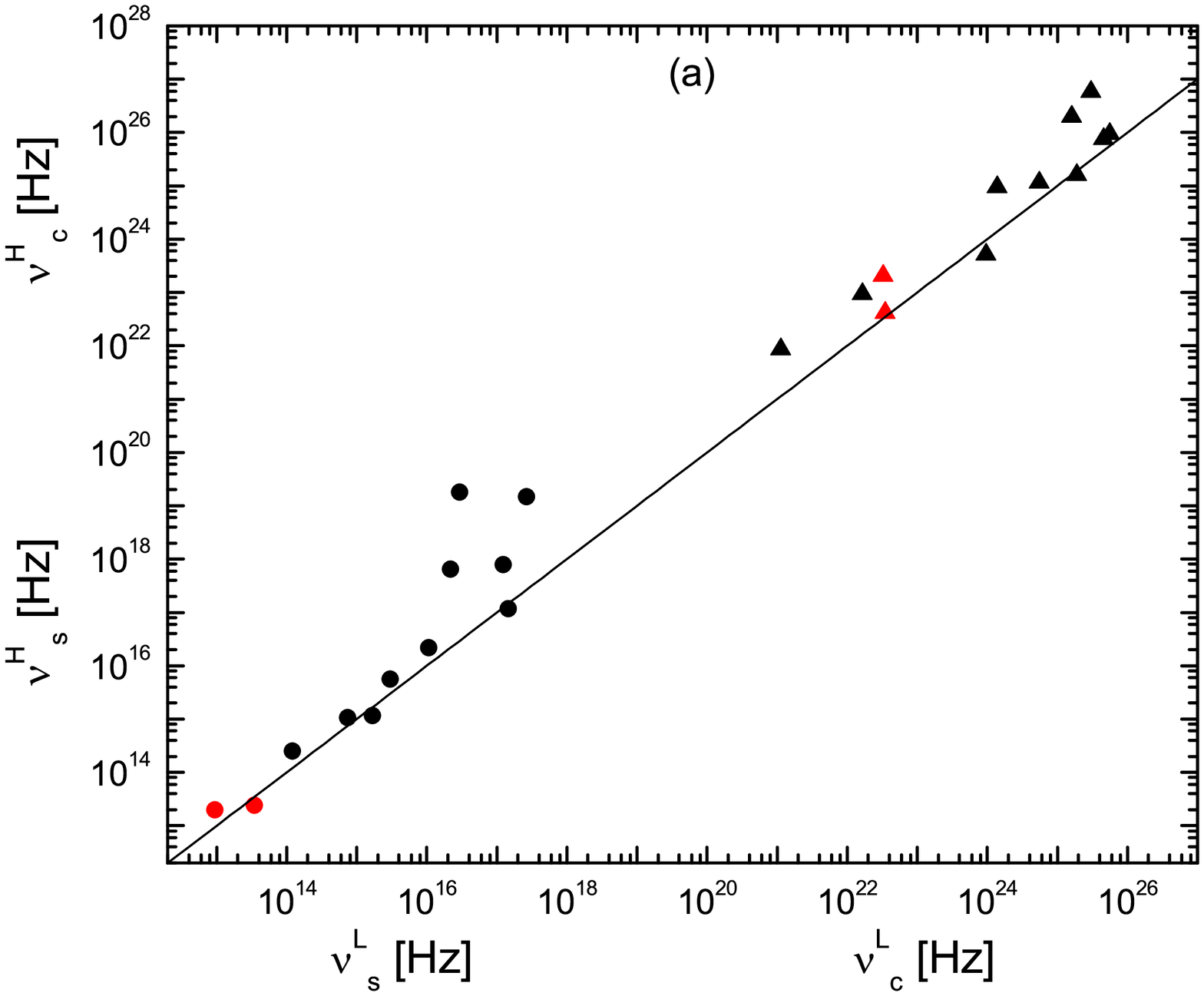}
\hfill
\includegraphics[angle=0,scale=0.27]{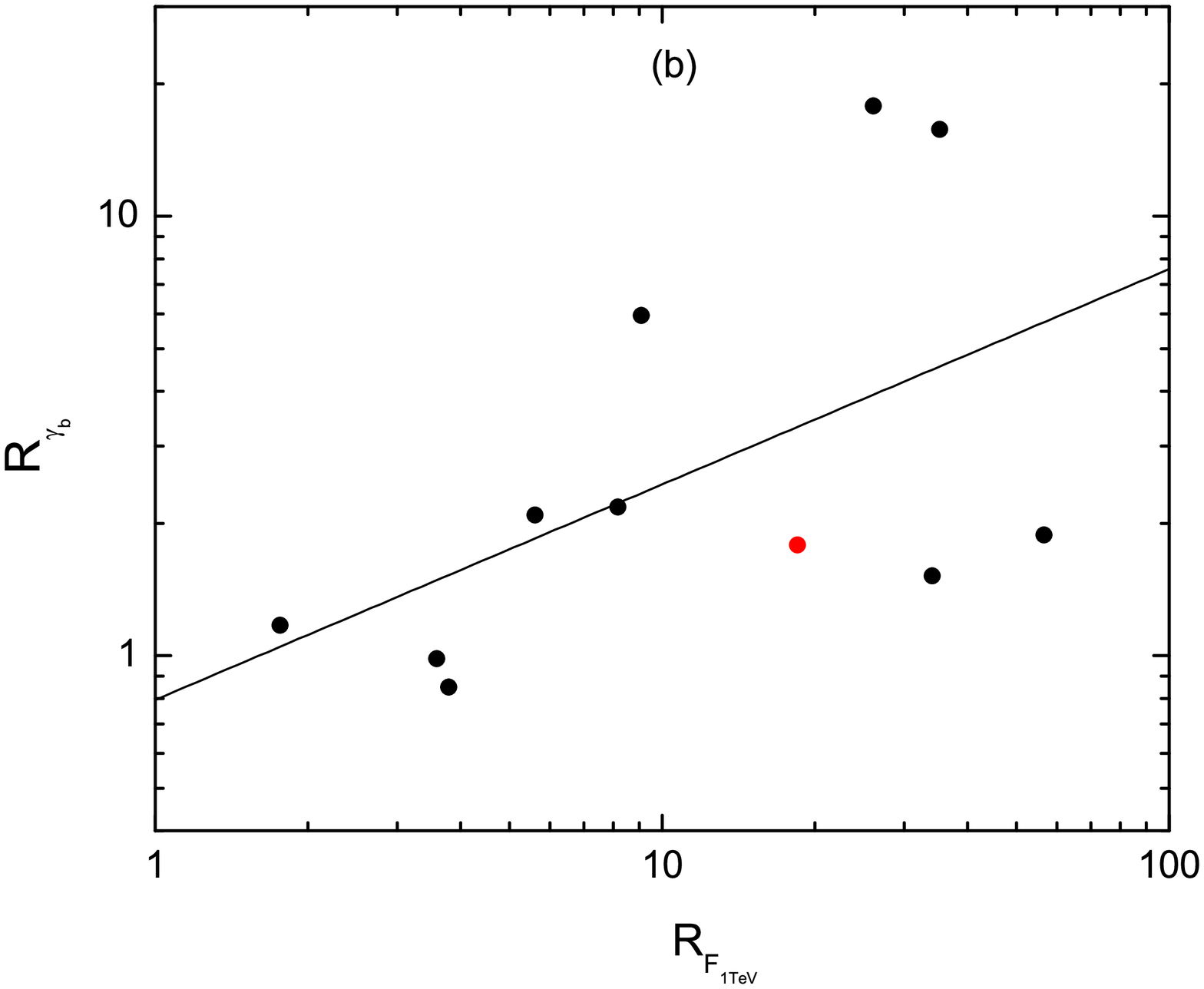}
\caption{{\em Panel a}---Comparison of the peak frequencies $\nu_{\rm s}$ ({\em circles}) and $\nu_{\rm c}$
({\em triangles}) between the high and low states. The {\em solid} line is the equality line. {\em Panel
b}---Ratio $R_{\gamma_{\rm b}}$ as a function of the ratio $R_{\rm 1\ TeV}$. The line is the best fitting line
$\log R_{\gamma_{\rm b}}=(-0.1\pm0.3)+(0.49\pm0.27)\log R_{\rm 1\ TeV}$. The red symbols are the data for the
FSRQs in our sample.}
\end{figure*}

\begin{figure*}
\includegraphics[angle=0,scale=0.27]{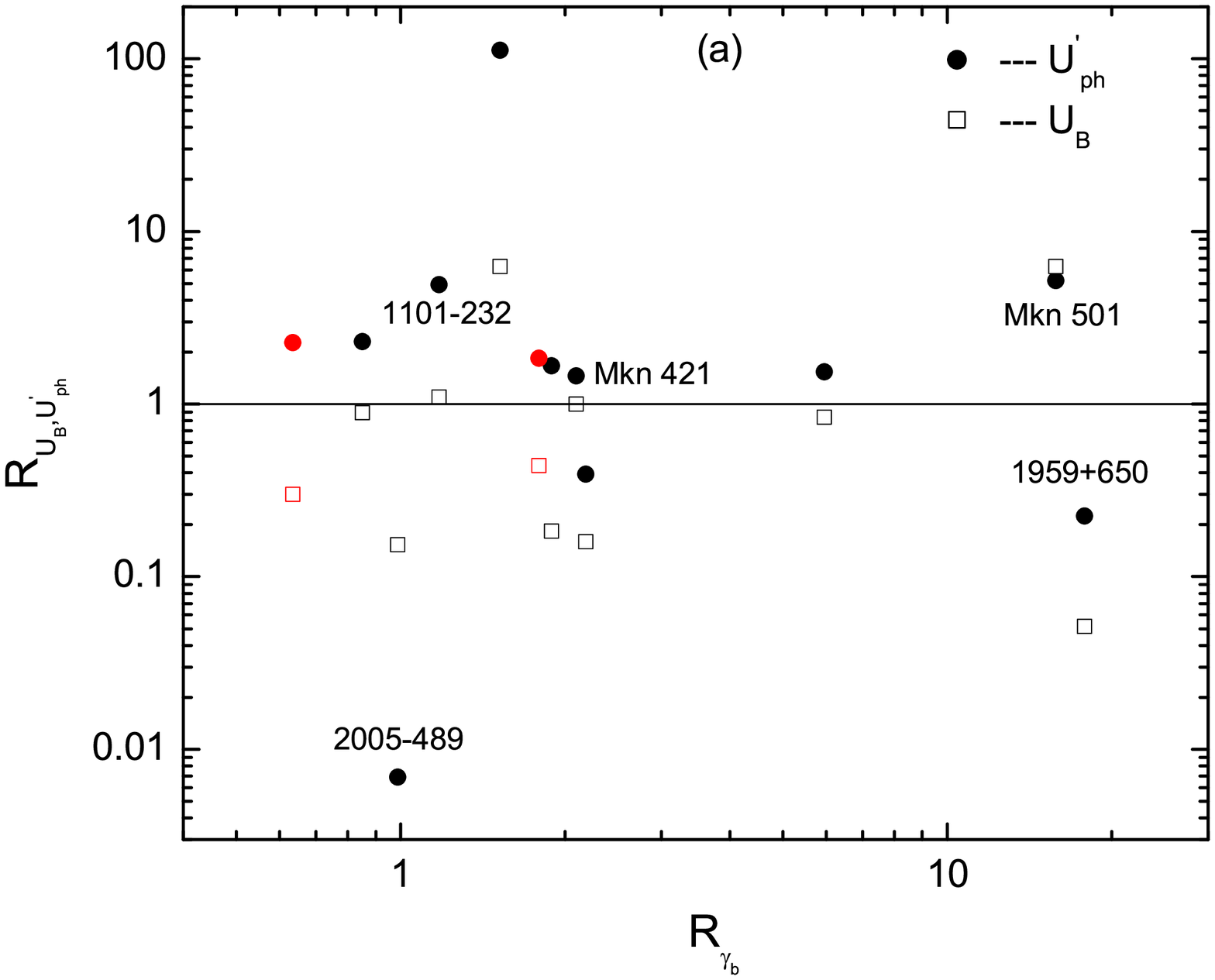}
\hfill
\includegraphics[angle=0,scale=0.27]{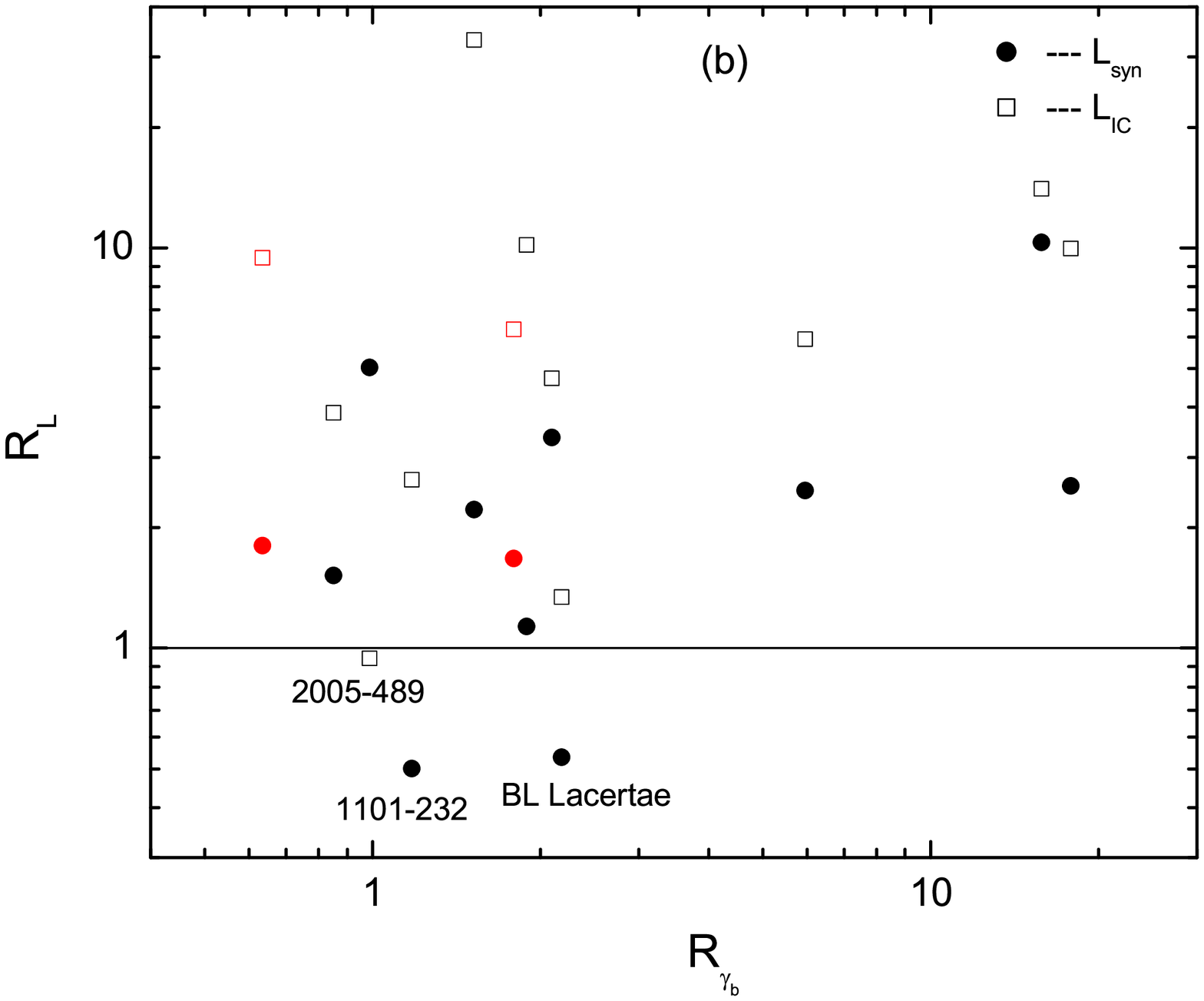}
\caption{The ratios of $U_{B}$, $U_{\rm ph}^{'}$, $L_{\rm syn}$, and $L_{\rm IC}$ as a function of the ratio of
$\gamma_{\rm b}$ for the high state to the low state. The red symbols are the data for the FSRQs in our sample.}
\end{figure*}

\begin{figure*}
\includegraphics[angle=0,scale=0.19]{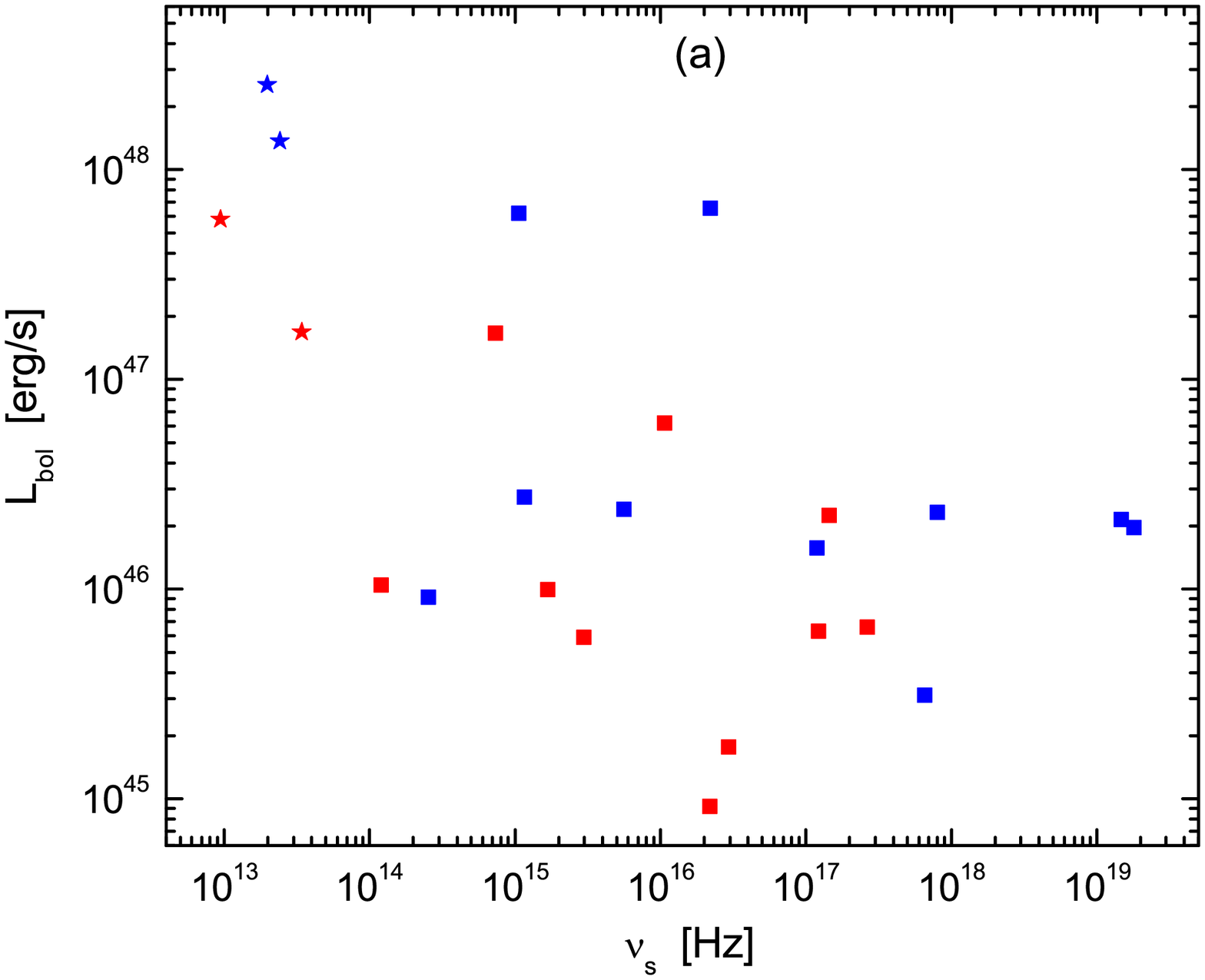}
\includegraphics[angle=0,scale=0.19]{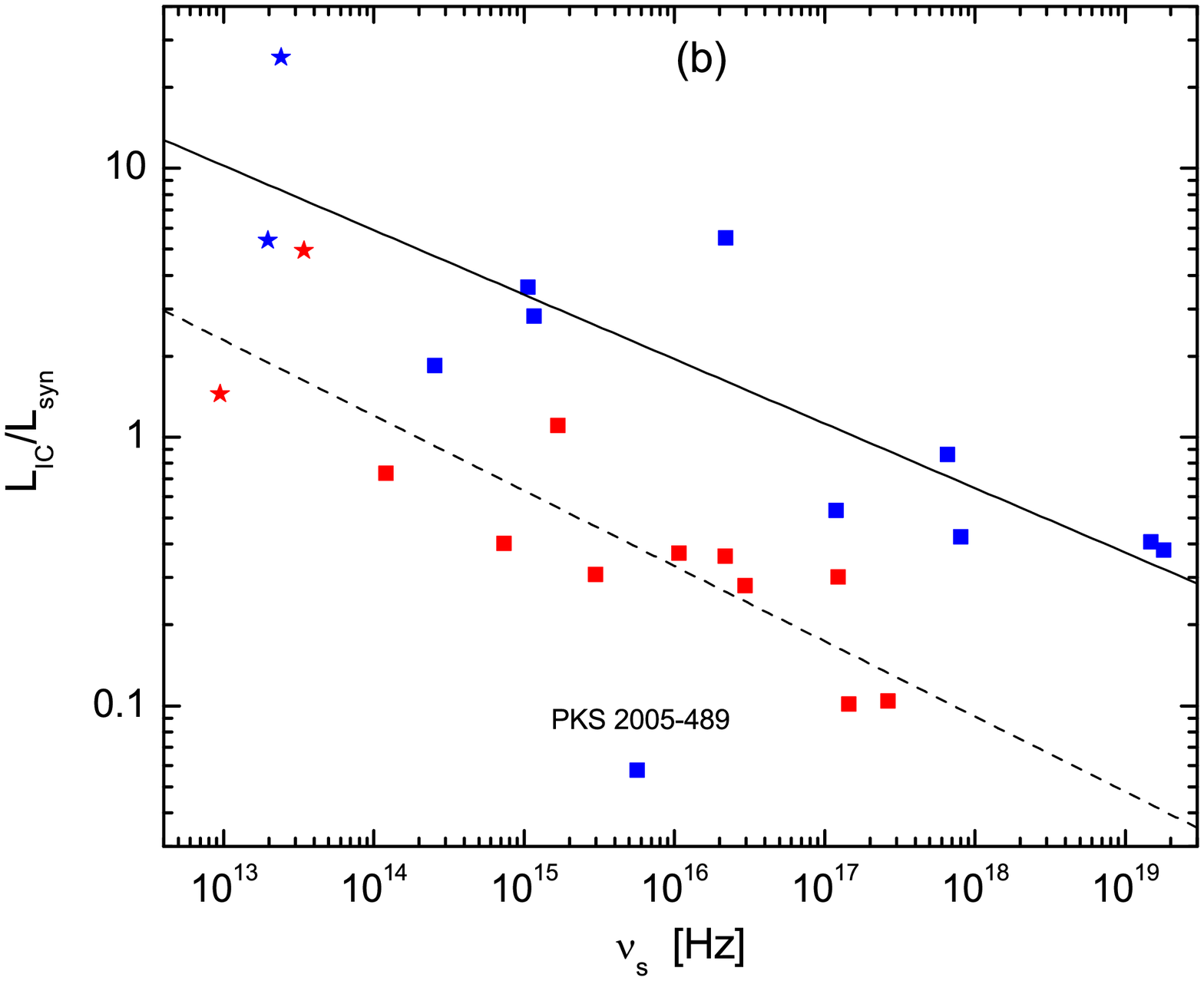}
\includegraphics[angle=0,scale=0.19]{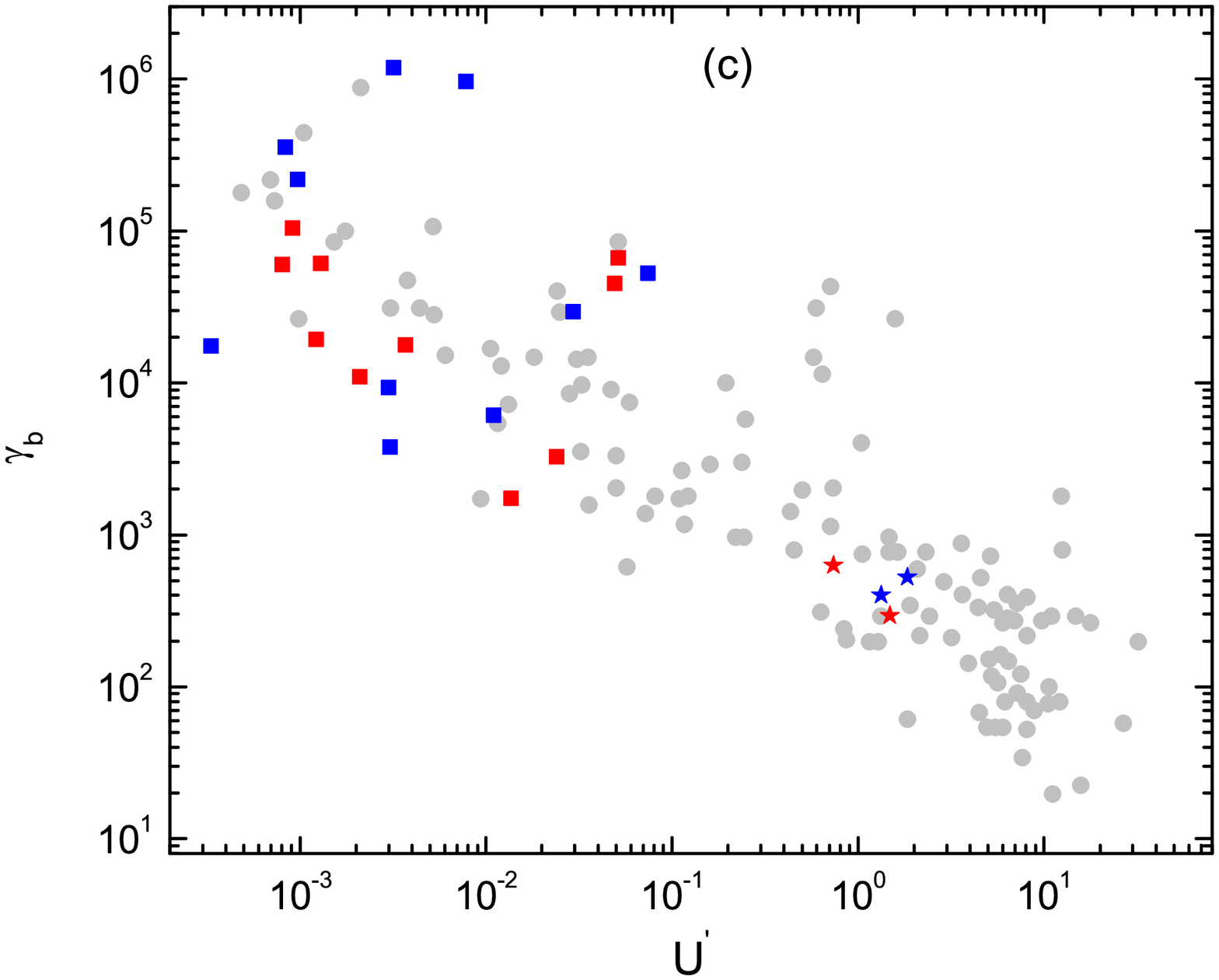}
\caption{{\em Panel a} and {\em Panel b}---Bolometric luminosity $L_{\rm bol}$ and the ratio $L_{\rm c}/L_{\rm
s}$ as a function of $\nu_{\rm s}$. The data of high and low states are marked with {\em blue and red} symbols,
respectively, and the stars are the data for two FSRQs. The best fit lines in {\em Panel b} are $\log (L_{\rm
c}/L_{\rm s})=(4.0\pm0.8)-(0.28\pm0.05)\log {\nu_{\rm s}}$ for the low state data ({\em dashed line}) and
$\log (L_{\rm c}/L_{\rm s})=(4.13\pm0.76)-(0.24\pm0.05)\log {\nu_{\rm s}}$ for the high state data ({\em solid
line}; excluding the source PKS 2005-489 in the high state). {\em Panel c}---$\gamma_{\rm b}$ as a function of
energy density $U^{'}$. The {\em gray circles} in {\em Panel c} are the sample data from Ghisellini et al.
(2010).}
\end{figure*}

\begin{figure*}
\includegraphics[angle=0,scale=0.19]{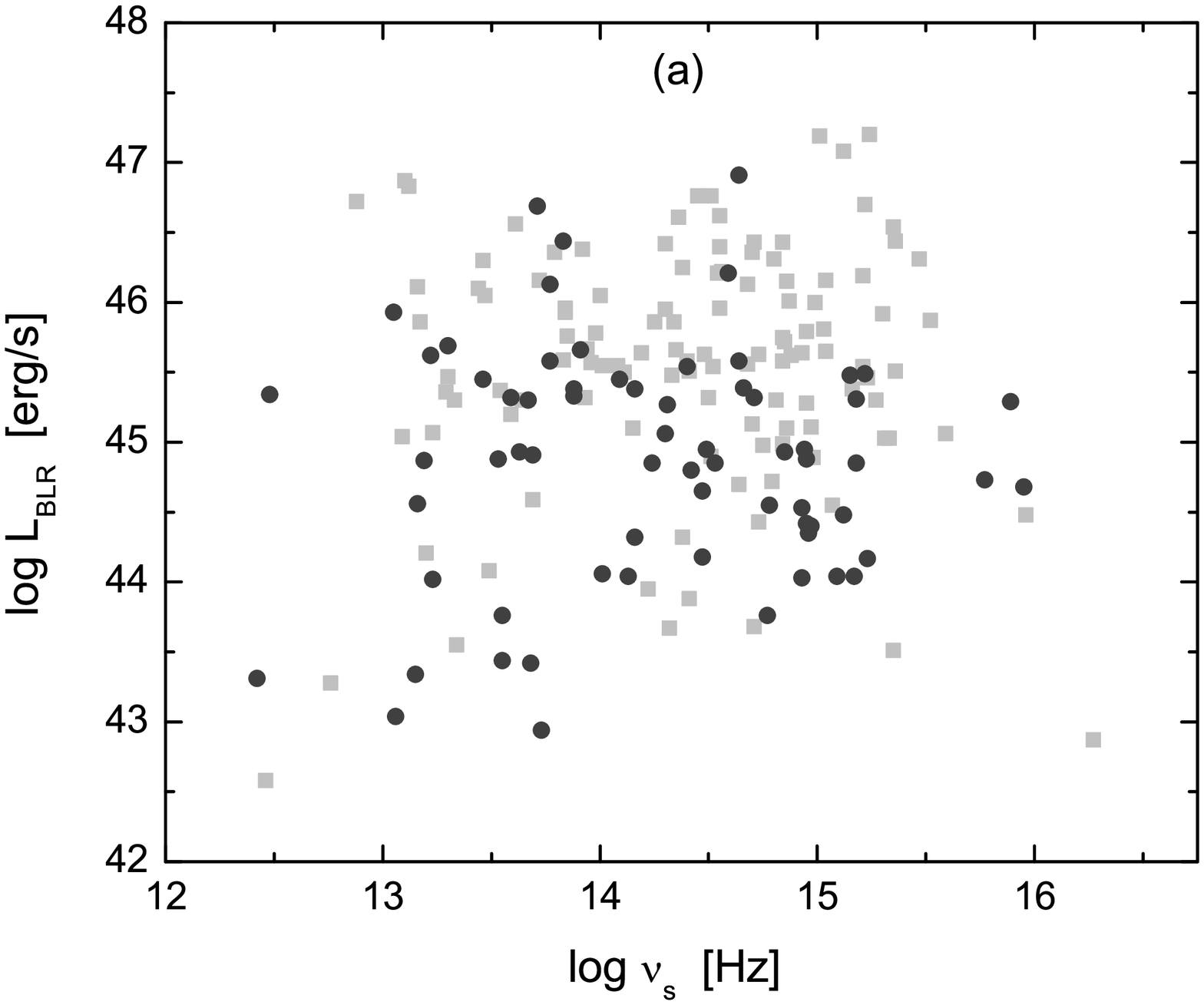}
\includegraphics[angle=0,scale=0.19]{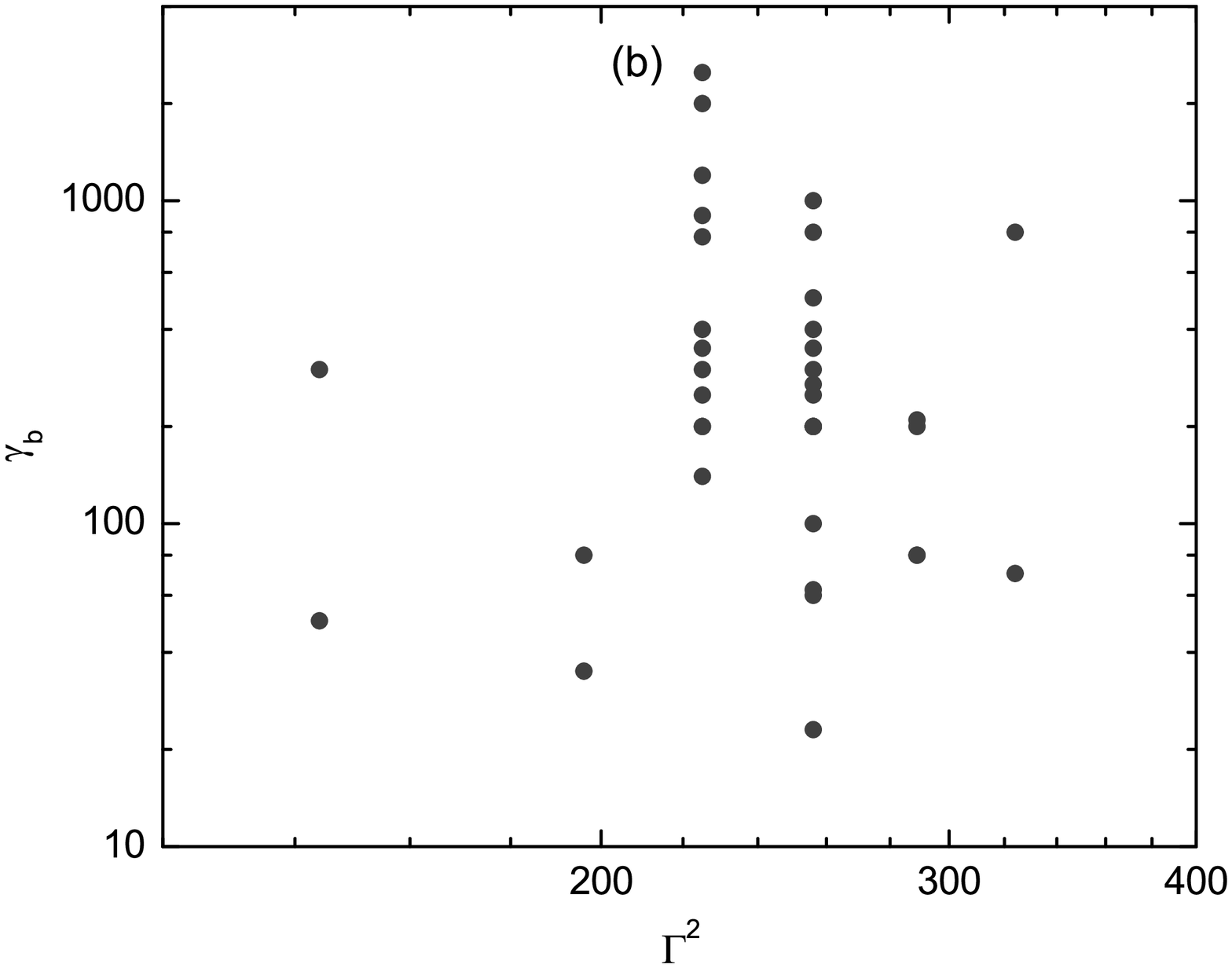}
\includegraphics[angle=0,scale=0.19]{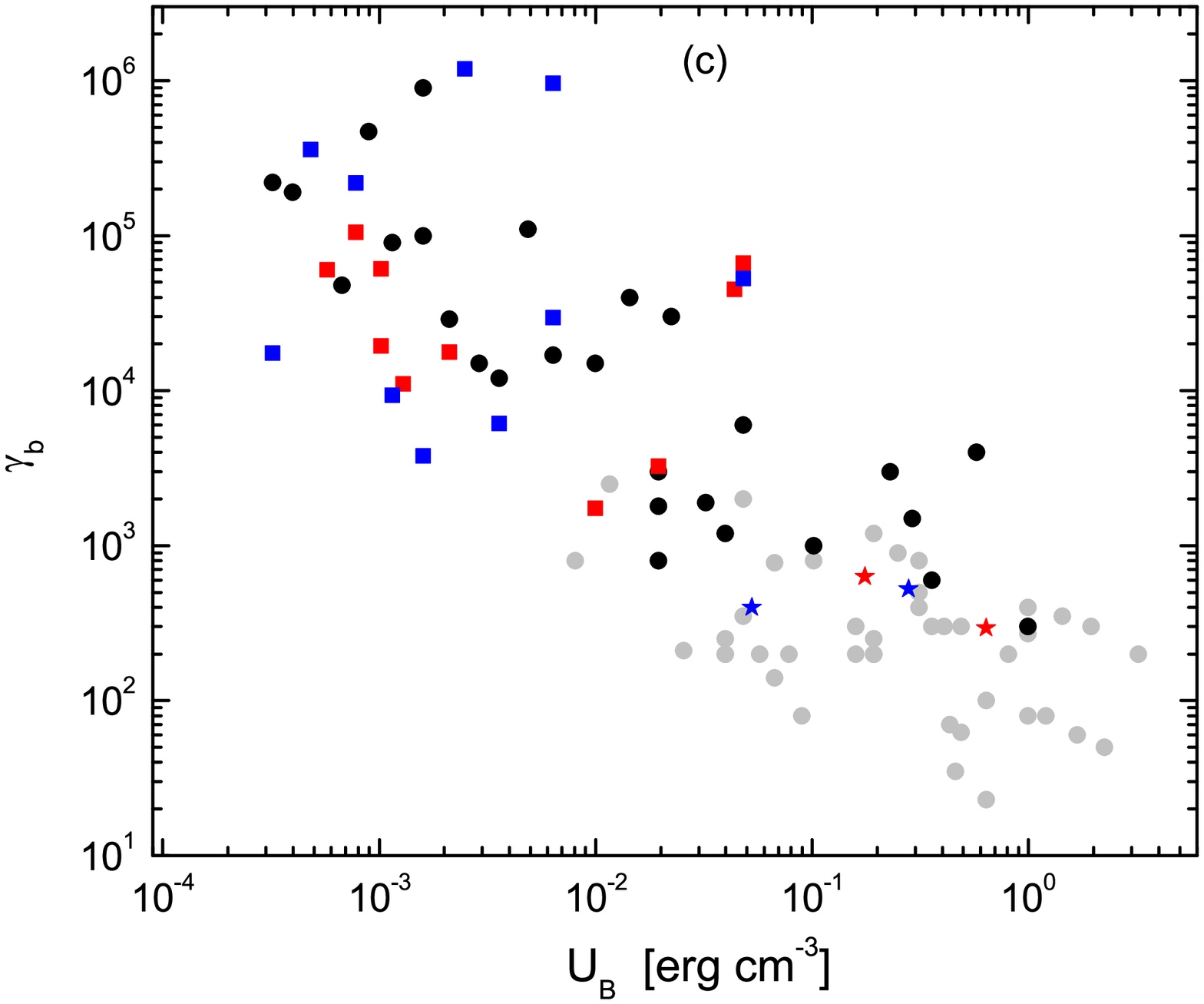}
\caption{{\em Panel a}---The luminosity of BLRs ($L_{\rm BLR}$)as a function of synchrotron radiation peak
frequency $\nu_{\rm s}$. The data are from Chen et al. (2009). The {\em back} and the {\em gray} symbols
indicate thermal-dominated and non-thermal-dominated FSRQs, respectively. {\em Panel b, c}---$\gamma_{\rm b}$ as
the functions of $\Gamma^{2}$ and $U_{B}$. The data marked as {\em circles} are from Celotti \& Ghisellini
(2008). In the {\em Panel c}, {\em black circles} and {\em light gray circles} indicate BL Lacs and FSRQs,
respectively, and the other symbols are the same as Fig.~4.}
\end{figure*}

\begin{figure*}
\includegraphics[angle=0,scale=0.19]{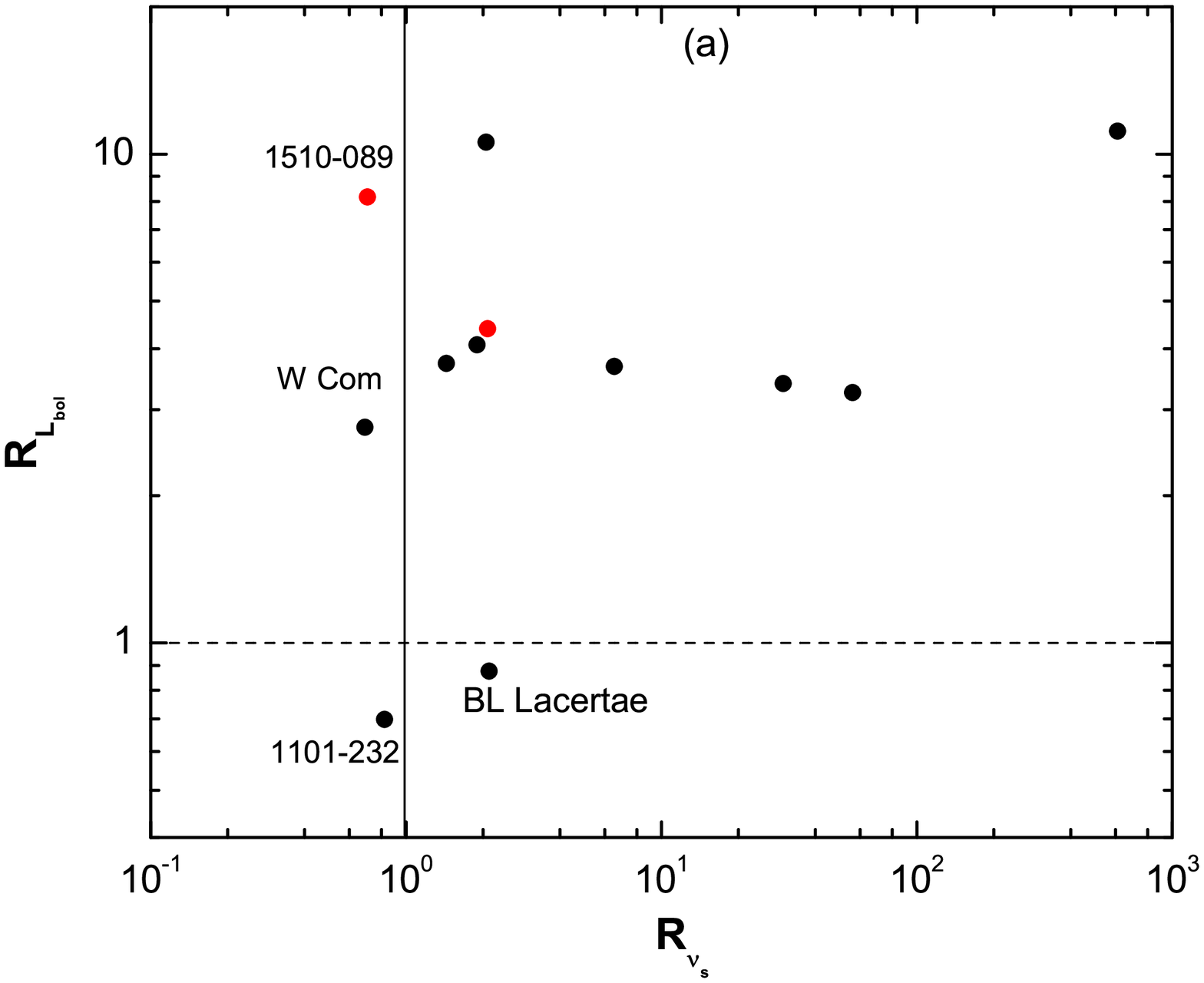}
\includegraphics[angle=0,scale=0.19]{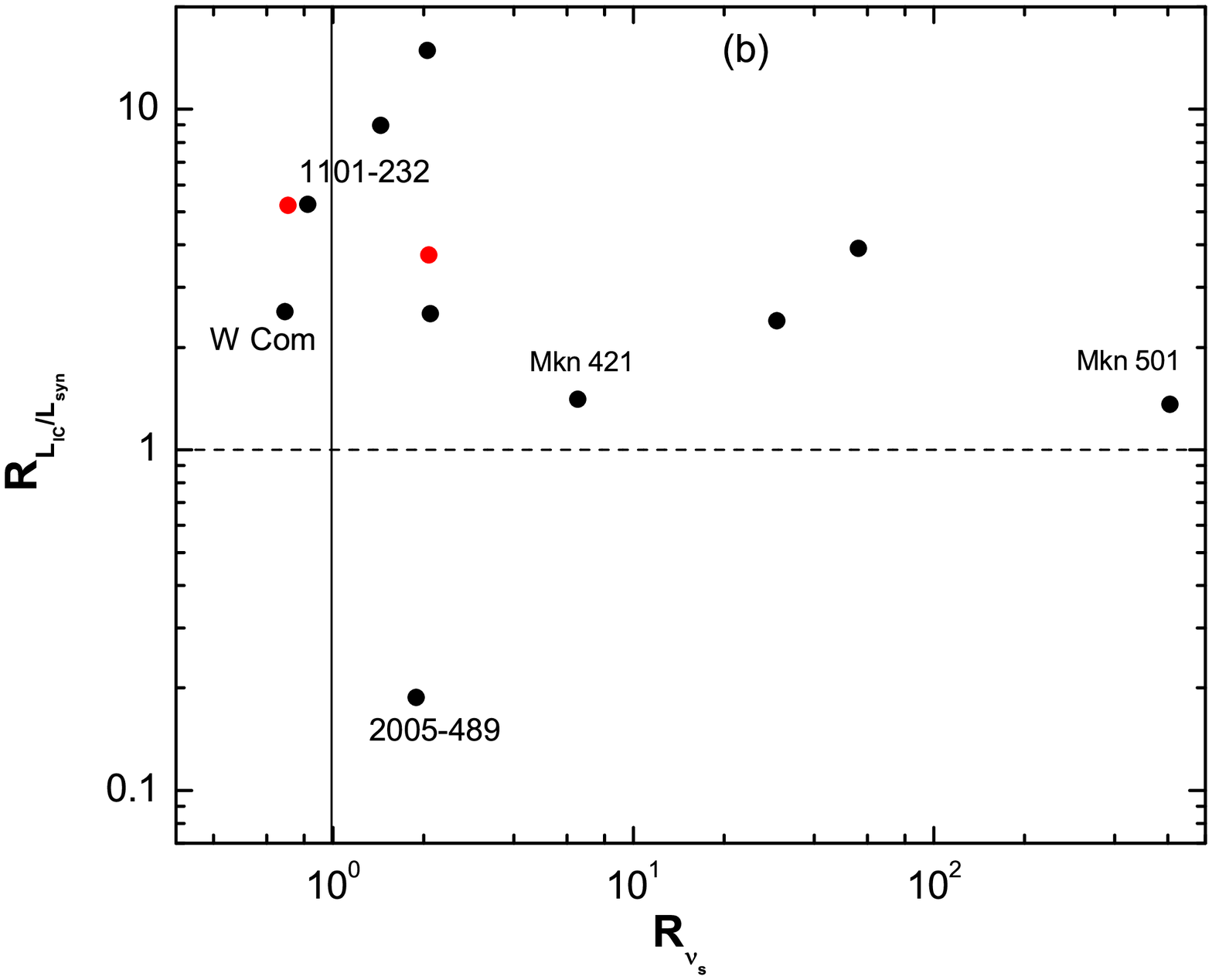}
\includegraphics[angle=0,scale=0.19]{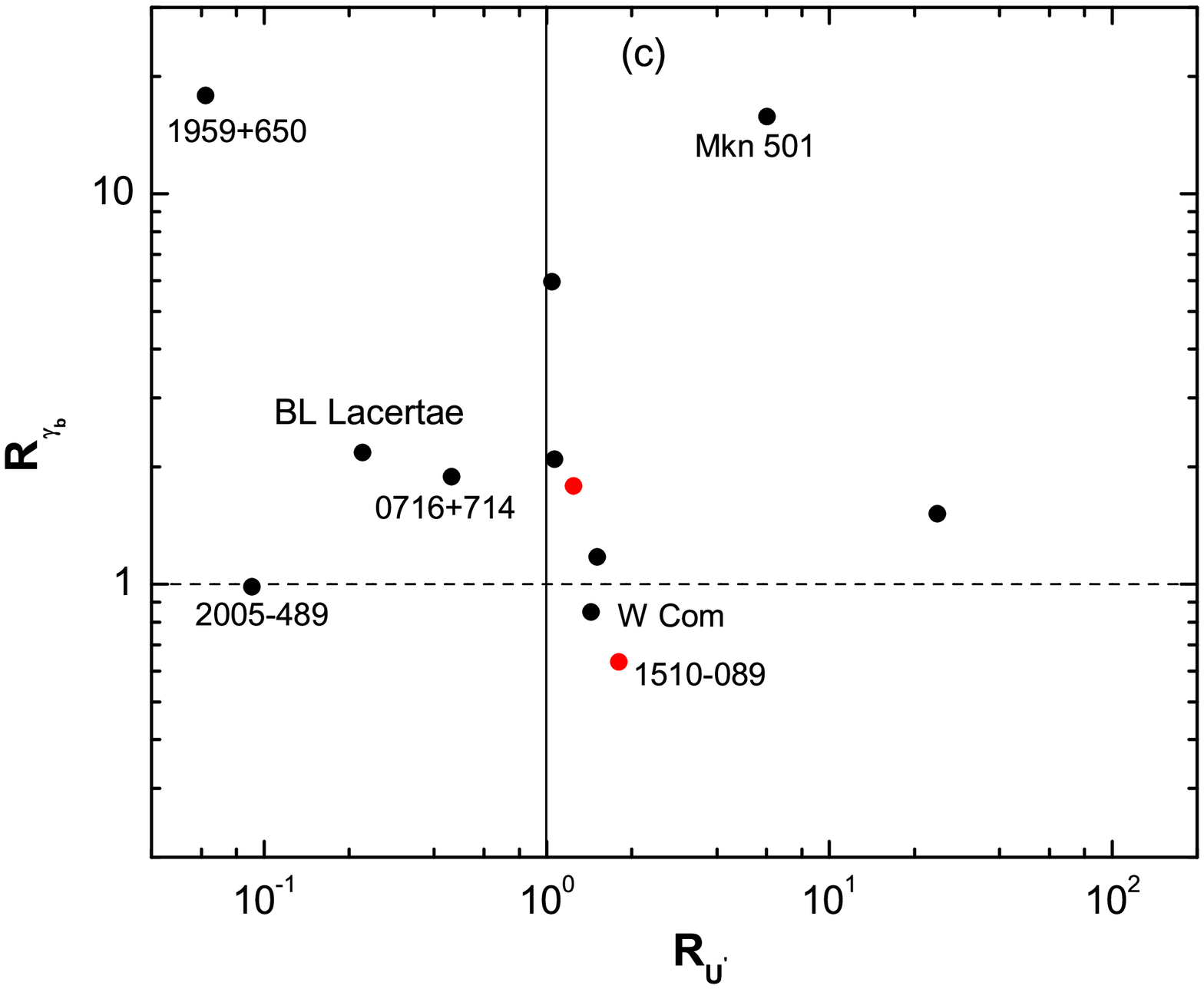}
\caption{{\em Panel a, b}---Ratios $R_{L_{\rm bol}}$ and $R_{L_{\rm c}/L_{\rm _s}}$ as a function of the ratio
$R_{\nu_{\rm s}}$ for high and low states. {\em Panel c}---Ratio $R_{\gamma_{\rm b}}$ as a function of the ratio
$R_{U^{'}}$. The symbols are the same as Fig.~2.}
\end{figure*}

\begin{figure*}
\includegraphics[width=4.in,height=2.in]{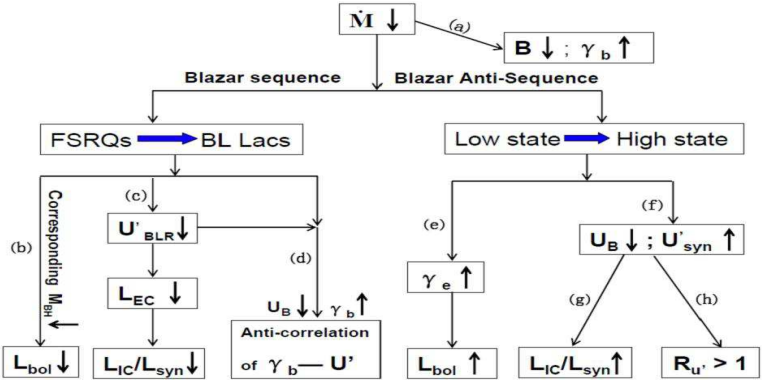}
\caption{Flow chart on the mechanism behind the blazar sequence for different types of sources and anti-sequence
for spectral variability of individual sources, both of which are driven by mass accretion rate $\dot M$. The
downward or upward arrows besides a parameter indicates decrease or increase of the parameter. Other arrows in
the flow chart describe the underlying causal relations. See the text for detailed descriptions.}
\end{figure*}

\end{document}